\documentclass[11pt]{article}
\usepackage{moriond,epsfig}
\usepackage{floatflt}
\usepackage{wrapfig}

\newcommand{\gevc} { {\rm GeV}/c}
\newcommand{\gevcc}{ {\rm GeV/}c^2}

\begin{document}
\vspace*{4cm}
\title{Multi-muon events at CDF}

\author{F.~Ptochos on behalf of the CDF collaboration}

\address{University of Cyprus, Physics Department, Nicosia, 1678 Cyprus}

\maketitle\abstracts{
 We report a study of multi-muon events produced at the Fermilab Tevatron
 collider and recorded by the CDF~II detector. In a data set acquired with
 a dedicated dimuon trigger and corresponding to an integrated luminosity
 of 2100 pb$^{-1}$, we isolate a significant sample of events in which at 
 least one of the identified muons has large impact parameter and is
 produced outside the beam pipe of radius 1.5 cm. We are unable to fully 
 account for the number and properties of the events through standard
 model processes in conjunction with our current understanding of the CDF~II
 detector, trigger and event reconstruction. Several topological and 
 kinematic properties of these events are also presented. 
 In contrast, the production cross section and kinematics of events in which
 both muon candidates are produced inside the beam pipe are successfully
 modeled by known QCD processes which include heavy flavor production.
 The presence of these anomalous multi-muon events offers a plausible resolution
 to long-standing inconsistencies related to $b\bar{b}$ production and decay.  
}
\section{Introduction}
 This study reports the observation of an  anomalous muon production in
 $p\bar{p}$ interactions at $\sqrt{s}=1.96$ TeV. The analysis was  motivated
 by the presence of several inconsistencies that affect or affected the
 $b\bar{b}$ production at the Tevatron:
 (a) the ratio of the observed  $b\bar{b}$ correlated production cross 
 section to the exact next-to-leading-order (NLO) QCD prediction~\cite{mnr}
 is $1.15 \pm 0.21 $ when  $b$ quarks are selected via secondary vertex 
 identification, whereas this ratio is found to be significantly larger
 than two when $b$ quarks are identified through their semileptonic 
 decays~\cite{bstatus}; (b) sequential semileptonic decays of single 
 $b$ quarks are considered to be the main source of dilepton events with 
 invariant mass smaller than that of a $b$ quark. However, the observed 
 invariant mass spectrum is not well modeled by the standard model (SM)
 simulation of this process~\cite{dilb}; and (c) the value of $\bar{\chi}$,
 the average time integrated mixing probability of $b$ flavored hadrons
 derived from the ratio of muon pairs from $b$ and $\bar{b}$ quarks
 semileptonic decays with opposite and same sign charge, is measured at
 hadron colliders to be larger than that measured by the LEP
 experiments~\cite{bmix,pdg}
 This analysis extends a recent study~\cite{bbxs} by the CDF collaboration 
 which has used a dimuon data sample to measure the correlated
 $\sigma_{b\rightarrow\mu,\bar{b}\rightarrow \mu}$ cross section.
 After briefly describing that study, it is shown that varying the
 dimuon selection criteria isolates a sizable, but unexpected background 
 that contains muons with an anomalous impact parameter~\cite{d0}
 distribution. Further investigation shows that a smaller fraction of these
 events also has anomalously large track and muon multiplicities. We are
 unable to account for the size and properties of these events in terms
 of known SM processes, even in conjunction with possible detector
 mismeasurement effects.

 The CDF~II detector~\cite{cdfdet} consists of a magnetic spectrometer,
 based on a 96-layer drift chamber, surrounded by electromagnetic and hadron
 calorimeters and muon detectors. Precision impact parameter and vertex
 determinations are provided by three slicon tracking devices collectively
 referred to in this report as the ``SVX". The SVX is composed of eight
 layers of silicon microstrip detectors ranging in radius from $1.5$ to
 $28$~cm in the pseudorapidity region $|\eta|<1$.
 \section{Study of the data sample composition}
 The study presented here, which is further detailed in Ref,\cite{a0disc}
 uses  the same data and Monte Carlo simulated samples, and the same analysis
 methods described in  Ref.~\cite{bbxs} We use events containing two central
 ($|\eta|<0.7$) muons, each with transverse momentum $p_T \geq 3 \; \gevc$,
 and with invariant mass larger than 5 $\gevcc$. In Ref,\cite{bbxs} the 
 value of  $\sigma_{b\rightarrow\mu,\bar{b}\rightarrow \mu}$ is determined 
 by fitting the impact parameter distribution of these primary muons
 with the expected shapes from all known sources. To ensure an accurate
 impact parameter determination,  Ref.\cite{bbxs}  uses a subset of 
 dimuon events in which each muon track is reconstructed in the SVX with
 hits in the two inner layers and in at least four of the inner six layers.  The data are nicely
 \begin{wrapfigure}[15]{l}{8.2cm}
 \vspace{-0.4cm}
 \includegraphics[width=8.0cm]{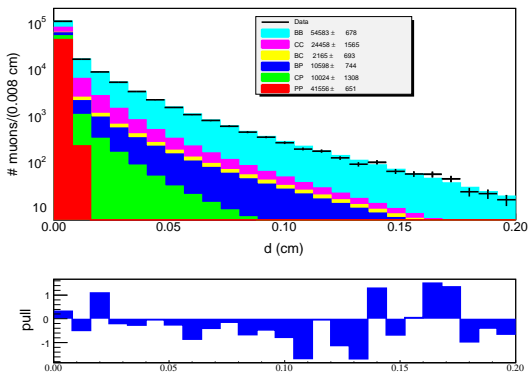}
 \vspace{-0.2cm}
 \caption[] {Impact parameter distribution of muons contributed
             by different physics processes.}
 \label{fig:fig01}
 \end{wrapfigure}
 described by a fit with contributions from the
 following QCD processes: semileptonic heavy flavor decays, prompt 
 quarkonia decays, Drell-Yan production, and instrumental backgrounds
 from hadrons mimicking the muon signal. Using the fit result, shown
 in Fig.~\ref{fig:fig01}, Ref.~\cite{bbxs} reports 
 $\sigma_{b\rightarrow\mu,\bar{b}\rightarrow \mu}= 1549 \pm 133$ pb
 for muons with $p_T \geq 3 \; \gevc$ and $|\eta| \leq 0.7$.

 This result is in good agreement with theoretical expectations as well as
 with analogous measurements that identify $b$ quarks via secondary vertex
 identification.\cite{ajets,shears} However, it is also substantially
 smaller than previous measurements of this cross section\cite{2mucdf,d0b2},
 and raises some  concern about the composition of the initial dimuon 
 sample prior to the SVX requirements. The tight SVX requirements used in
 Ref.\cite{bbxs} select events in which both muons arise from parent 
 particles that have decayed within a distance of $\simeq 1.5$ cm from the
 $p\bar{p}$ interaction primary vertex in the plane transverse to the beam 
 line. Using Monte Carlo simulations, we estimate that approximately 96\% 
 of the dimuon events contributed by known QCD processes satisfy this latter
 condition. Since the events selected in~\cite{bbxs} are well described by
 known QCD processes, we can independently estimate the efficiency of the
 tight SVX requirements. Using control samples of data from various sources
 and the sample composition determined by the fit to the muon impact parameter
 distribution, we estimate that ($24.4\pm 0.2$)\% of the initial sample should
 survive the tight SVX requirements, whereas only ($19.30\pm0.04$)\% actually
 do. This suggests the presence of an additional background that has been
 suppressed when making the tight SVX requirements. The size of this 
 unexpected dimuon source is estimated as the difference of the total 
 number of dimuon events, prior to any SVX requirements, and the expected
 contribution from the known QCD sources. This latter contribution is
 estimated as the number of events surviving the tight SVX requirements
 divided by the efficiency of that selection. In a data set corresponding 
 to an integrated luminosity of 742 pb$^{-1}$,  143743 dimuon events survive
 the tight SVX cuts. Dividing this number by the $24.4\%$ efficiency of the 
 tight SVX selection criteria we expect $589111\pm 4829$ QCD  events to 
 contribute to the initial sample whereas 743006 are observed. The difference,
 $153895\pm 4829$ events, is comparable in magnitude to the expected dimuon
 contribution from $b \bar{b}$ production, $221564\pm 11615$. This estimate
 assumes the unexpected source of dimuon events is completely rejected by
 the tight SVX requirements.

 Most CDF analyses use a set of SVX criteria, referred in the following as 
 standard SVX, in which tracks are required to have hits in at least three
 of the eight SVX layers. This standard SVX selection accepts muons from
 parent particles with decay lengths as long as $10.6$~cm. Applying the
 standard SVX selection reduces the estimated size of the unknown dimuon 
 source by a factor of two, whereas $88$\% of the known QCD contribution
 is expected to survive.

 A summary of the estimates of the size of this unexpected source of dimuon 
 events, whimsically called ghost events, for various sets of SVX criteria 
 is shown in Table~\ref{tab:tab01}. In this table and throughout this report
 the expected contribution from known QCD sources, referred to as QCD 
 contribution, will be estimated from the sample of dimuons surviving the
 tight SVX requirements and properly accounting for the relevant SVX 
 efficiencies using the sample composition from the
 \begin{wraptable}[15]{l}{0.65\textwidth}
 \vspace{-0.4cm}
 \caption[]{Number of events that pass different SVX requirements. 
            Dimuons are also split into pairs with opposite ($OS$)
            and same ($SS$) sign charge.}
 \vspace{0.3cm}
 \begin{tabular}{|lccc|}
 \hline
  Type       &  No SVX            & Tight SVX   & Standard SVX      \\
  Total      &  743006            & 143743      & 590970            \\
  Total $OS$ &                    &  98218      & 392020            \\
  Total $SS$ &                    &  45525      & 198950            \\
  QCD        & 589111 $\pm$ 4829  & 143743      & 518417 $\pm$ 7264 \\   
  QCD   $OS$ &                    &  98218      & 354228 $\pm$ 4963 \\
  QCD   $SS$ &                    &  45525      & 164188 $\pm$ 2301 \\
  Ghost      & 153895 $\pm$ 4829  &   0         & 72553  $\pm$ 7264 \\
  Ghost $OS$ &                    &   0         & 37792  $\pm$ 4963 \\
  Ghost $SS$ &                    &   0         & 34762  $\pm$ 2301 \\ 
 \hline
 \end{tabular}
 \label{tab:tab01}
 \end{wraptable}
 fits of Ref.\cite{bbxs} We elect to follow this approach since the tight SVX sample provides a well
 understood sample.\cite{bbxs} The ghost contribution will always be
 estimated from the total number of events observed in the data after
 subtracting the expected QCD contribution. Table~\ref{tab:tab01} shows also 
 the event yields separately for the subset of events in which the dimuons
 have opposite-sign ($OS$) and same-sign ($SS$) charge. The ratio of OS to SS 
 dimuons is approximately 2:1 for QCD processes but is approximately 1:1
 for the ghost contribution.

 At this stage it is worth commenting further on the set of inconsistencies
 related to $b\bar{b}$ production and decay mentioned above. The general
 observation is that the measured $\sigma_{b\rightarrow\mu,\bar{b}\rightarrow \mu}$
 increases as the SVX requirements are made looser and is almost a factor of
 two larger than that measured in Ref.\cite{bbxs} when no SVX requirements are 
 made.\cite{d0b2} As mentioned above, the magnitude of the ghost contribution
 is comparable to the $b\bar{b}$ contribution when no SVX selection is made
 and in combination would account for the measurement reported in
 Ref.~\cite{d0b2} Similarly, for the standard SVX criteria, the magnitude of
 the ghost contribution, when added to the expected $b\bar{b}$ contribution of
 $194976 \pm 10221$ events, coincides with the cross section measurement 
 reported in Ref.\cite{2mucdf} and the $\bar{\chi}$ value reported in
 Ref.\cite {bmix} since these measurements use similar sets of silicon criteria.
 Moreover, as demonstrated in\cite{a0disc}, when applying the tight SVX 
 criteria to initial muons, the invariant mass spectrum of combinations of an
 initial muon with an additional accompanying muon is well described by known
 QCD sources and is dominated by sequential semileptonic heavy flavor decays. 
 In contrast, without any SVX requirement the invariant mass spectrum cannot 
 be modeled with the SM simulation and the inconsistencies at low invariant
 mass reported in\cite{dilb} are reproduced. Thus, this unknown source of
 dimuon events seems to offer a plausible resolution to these long-standing
 inconsistencies related to $b\bar{b}$ production and decay. The remainder
 of this paper is dedicated to a further exploration of these events.

 The nature of the anomalous events can be characterized by four main features.
 The impact parameter distribution of the initial muon pair cannot be readily
 understood in terms of known SM processes. In small angular cones around the
 initial muons the rate of additional muons is significantly higher than that
 expected from SM processes. The invariant mass of the initial and additional
 muons looks different from that expected from sequential semileptonic decays
 of heavy flavor hadrons. The impact parameter distribution of the additional
 muons has the same anomalous behavior as the initial muons. We will discuss
 these features in turn.

\begin{wrapfigure}[23]{l}{7.0cm}
 \vspace{-0.5cm}
 \includegraphics[width=6.8cm]{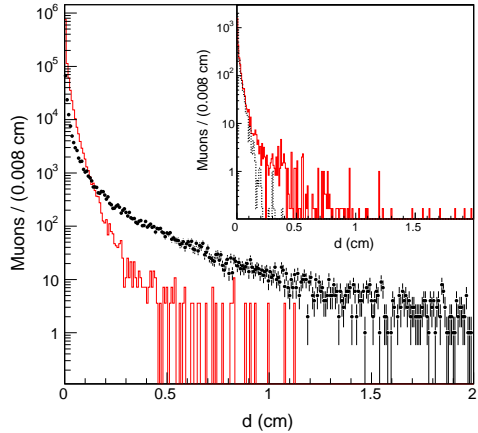}
 \vspace{-0.3cm}
 \caption[] {Impact parameter distribution of muons contributed by ghost
             ($\bullet$) and QCD (histogram) events. Muon tracks are selected
             with the standard SVX requirements. The detector resolution is
             $\simeq 30 \; \mu$m. The insert shows the distribution of 
             simulated muons (histogram) that pass the same analysis selection
             as the data and arise from in-flight-decays of pions and
             kaons produced in a QCD heavy flavor simulation. The dashed
             histogram shows the impact parameter of the parent hadrons. }
 \label{fig:fig02}
 \end{wrapfigure}

 As shown in Fig.~\ref{fig:fig02}, muons due to ghost events have an impact parameter distribution
 that is completely different from that of muons due to QCD events.

 A number of potential background sources have been evaluated. The one expected
 to contribute significantly arises from in-flight-decays of pions
 and kaons. Based upon a generic QCD simulation, we predict a contribution
 of 57000 events,\cite{a0disc} 44\% and 8\% of which pass the standard and
 tight SVX selection, respectively. The uncertainty of this prediction is
 difficult to assess, but, as shown by the insert in Fig.~\ref{fig:fig02}, 
 in-flight decays alone cannot account for the shape of the muon impact
 parameter distribution in ghost events. A minor contribution of $K^0_S$
 and hyperon decays in which the punchthrough of a hadronic prong mimics
 a muons signal has been also investigated.\cite{a0disc}
 Secondary interactions in the tracking volume are also possible candidates,
 and more difficult to quantify. The possibility of instrumental effects,
 trigger and reconstruction biases have been investigated in detail in
 Ref.\cite{a0disc} For example, we have verified the soundness of large
 impact parameter tracks by measuring the lifetime of  $K^0_S$ decays
 reconstructed in the same data set used for this analysis.
\section{Events with additional muons}
 We search QCD and ghost events that contain a pair of initial muons that 
 pass our analysis selection (without any SVX requirement) for additional 
 muons with $p_T \geq 2 \; \gevc$ and $|\eta|\leq 1.1$. We have the following
 motivations: (a) events acquired because of in-flight decays or secondary 
 interactions are not expected to contain an appreciable number of additional
 muons; (b) QCD events that might appear in the ghost sample because of 
 not-yet-understood detector malfunctions should not contain more additional
 leptons than QCD events with well reconstructed initial dimuons; and
 (c) we want to investigate if the anomaly reported in Ref.\cite{dilb} is 
 also related to the presence of the unexpected background. According to the
 simulation,\cite{a0disc} additional muons arise from sequential decays of
 single $b$ hadrons. In addition, one expects a contribution due to hadrons
 mimicking the muon signal. In the data, 9.7\% of the dimuon events contain
 an additional muon (71835 out of 743006 events). The contribution of events
 without heavy flavor, such as all conventional sources of ghost
 events mentioned above, is depressed by the request of an additional muon.
 For example, in events containing a $\Upsilon(1S)$ or $K^0_S$ candidate
 and are included in the dimuon sample, the probability of finding an 
 additional muon is ($0.90 \pm 0.01$)\% and ($1.7 \pm 0.8$)\%, respectively.
 However, the efficiency of the tight SVX selection in  dimuon events that
 contain additional muons drops from $0.1930 \pm 0.0004$ to $0.166 \pm 0.001$.
 This observation anticipates that a fraction of ghost events contains
 more additional muons than QCD data.

 This paragraph summarizes a detailed study of the rate and kinematic
 properties of events that contain at least three muons reported in
 Ref.\cite{a0disc} This study uses  a data set of larger integrated 
 luminosity that corresponds to $1131090\pm 9271$ QCD and
 $295481 \pm 9271$ ghost events. Reference\cite{a0disc} shows that the
 rate and kinematics of three-muon combinations are correctly modeled
 by the QCD simulation only if the two initial muons are selected with
 the tight SVX requirement. Muon pairs due to $b$ sequential decays peak
 at small invariant masses and small 
 \begin{wrapfigure}[16]{l}{6.0cm}
 \vspace{-0.4cm}
 \includegraphics[width=5.8cm]{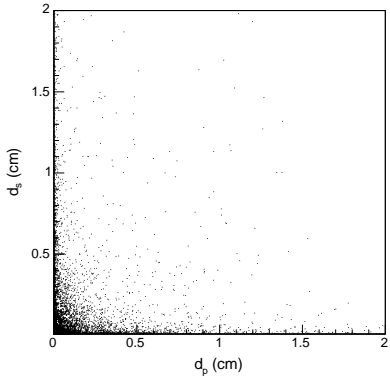}
 \vspace{-0.3cm}
 \caption[]{Two-dimensional distribution of the impact parameter of an
            initial muon, $d_{0p}$, versus that, $d_{0s}$, of additional
            muons in ghost events. Muons are selected with standard SVX 
            requirements.}
 \label{fig:fig03}
 \end{wrapfigure}
 opening angles. The distributions
 of analogous pairs in the unexpected background have a quite similar
 behaviour. However, combinations of initial and additional muons in
 ghost events have a smaller opening angle and a smaller invariant
 mass than those from sequential $b$ decays.\cite{a0disc}
 Therefore, the study of ghost events is further restricted to muons
 and tracks contained in a cone of angle $\theta \leq 36.8^\circ$
 ($\cos \theta\geq 0.8$) around the direction of each initial muon.
 As reported in Ref.,\cite{a0disc} less than half of the OS and SS muon
 combinations in ghost events can be accounted for by fake muons, and ghost
 events are shown to contain a fraction of additional real muons (9.4\%)
 that is four times larger than that of QCD events (2.1\%).
 Reference\cite{a0disc} investigates at length the possibility that
 the predicted rate of fake muons is underestimated. The fraction of
 additional real muons in QCD and ghost
 \begin{wrapfigure}[17]{l}{6.0cm}
 \vspace{-0.4cm}
 \includegraphics[width=5.8cm]{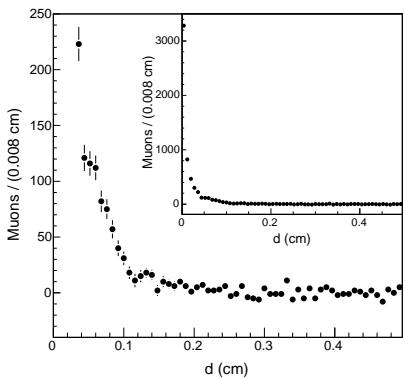}
 \vspace{-0.3cm}
 \caption[]{Exploded impact parameter distribution of additional muons in
            QCD events. The entire distribution is shown in the insert.
            Muons are selected without any SVX requirements.}
 \label{fig:fig04}
 \end{wrapfigure}
 events is verified by selecting additional muons with $p_T \geq 3\;\gevc$
 and $|\eta| \leq 0.7$. In this case, because of the larger number of
 interaction lengths traversed by hadronic tracks, the fake rate is
 negligible.\cite{bbxs} In this study the muon detector acceptance is
 reduced by a factor of five but the rate of such additional muons is
 ($0.40 \pm 0.01$)\% in QCD and $(1.64 \pm 0.08)$\% in ghost events.

 Figure~\ref{fig:fig03} shows the two-dimensional distribution of the impact
 parameter of an initial muon versus that of all additional muons in a 
 $\cos \theta \geq 0.8$ cone around its direction. The impact parameter
 distribution of the additional muons is found to be as anomalous as that of
 primary muons. However, the impact parameter of the additional and initial
 muons are weakly correlated (the correlation factor is $\rho_{d_{0p}d_{0s}}=0.03$).

 For comparison, Fig.~\ref{fig:fig04} shows that the impact parameter
 distribution of additional muons in QCD events is not anomalous at all.

 It is difficult to reconcile the rate and characteristics
 of these anomalous events with expectations from known SM sources. Although 
 one can never rule out the possibility that these data could be at least
 partially explained by detector effects not presently understood,
 we will present some additional properties of the ghost sample.

\begin{wrapfigure}[14]{l}{10.0cm}
 \vspace{-0.4cm}
 \includegraphics[width=10cm]{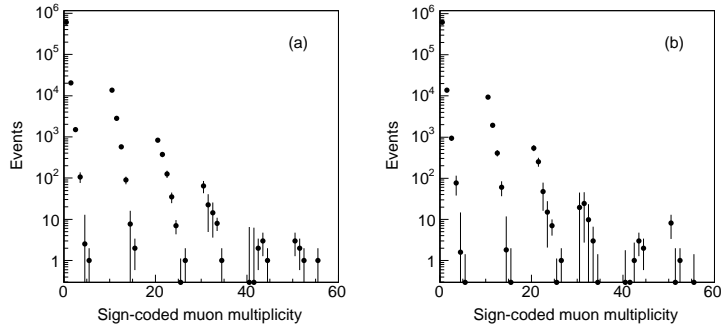}
 \vspace{-0.3cm}
 \caption[]{Multiplicity distribution  of additional muons found in a
            $\cos \theta \geq 0.8$ cone around the direction of a primary
            muon before (a) and after (b) correcting for the fake muon
            contribution. An additional muon increases the multiplicity by
            1 when it has opposite and by 10 when it has same sign charge as
            the initial muon.} 
 \label{fig:fig05}
 \end{wrapfigure}

 Figure~\ref{fig:fig05}~(a) shows the distribution of the number of muons
 found in a $\cos\theta \geq 0.8$ cone around a primary muon in ghost
 events. In the plot, an additional muon increases the multiplicity by 1
 when of opposite and by 10 when of same sign charge as the initial muon.
 Leaving aside the case in which no additional muons are found, it is 
 interesting to note that an increase of one unit in the muon multiplicity
 corresponds in average to a population decrease of approximately a factor
 of seven. This factor is very close to the inverse of the
 $\tau \rightarrow \mu$ branching fraction (0.174) multiplied by the 83\%
 efficiency of the muon detector, and makes it hard to resist the
 interpretation that these muons arise from $\tau$ decays with
 a kinematic acceptance close to unity. The multiplicity distribution
 corrected for the fake muon contribution\cite{a0disc} is shown in 
 Fig.~\ref{fig:fig05}~(b). The fake contribution is evaluated on a
 track-by-track basis using the probability that pions from $D^0$ mesons from
 $B$ hadron decays mimic a muon signal. Unfortunately, the multiplicity 
 distribution of muons and tracks contained in a $36.8^{\circ}$ cone around
 the direction of such $D^0$ mesons does not have the high multiplicity tail 
 of ghost events. In the $D^0$ control sample, we do not observe any
 dependence of the fake rate on the track and muon multiplicity, but we also
 cannot rule out a drastic increase of the fake probability per track in
 events with multiplicities much larger than those of QCD standard processes.
 A study based on higher quality muons\cite{a0disc} does not show any
 evidence of that being the case.\\
 \section{Conclusions}
 We report the observation of anomalous muon production in $p\bar{p}$
 collisions at $\sqrt{s}=1.96\, \rm TeV$. This unknown source of dimuon events
 seems to offer a plausible resolution to long-standing inconsistencies related
 to $b\bar{b}$ production and decay. A significant fraction of these events 
 has features that cannot be explained with our current understanding of the
 CDF~II detector, trigger and event reconstruction. 


\begin{thebibliography}{99}
\bibitem{d0}      The impact paramter is defined as the distance of closest
                   approach of a track to the primary event vertex in the 
                   transverse plane with respect to the beamline.
\end{thebibliography}
\end{document}